\documentclass{icrc2009}
\usepackage{graphicx}   % for including figures
\usepackage[caption=false]{caption}    % for captions
\usepackage[font=footnotesize]{subfig} % subfig.sty for a double column floating figure using two subfigures
\usepackage{fixltx2e} 
\usepackage{url}

\newcommand{\shorttitle}[1]%
{\markboth{Procetedings of the 31\MakeLowercase{$^{st}$} ICRC, {\L}\'{o}d\'{z} 2009}{#1} }
\newcommand{\etal}{\MakeLowercase{\textit{et al. }}} % "et al."
\begin{document}
\title{Search for neutrinos from Gamma-Ray Bursts with the Baikal neutrino telescope NT200}

\author{\IEEEauthorblockN{A. Avrorin\IEEEauthorrefmark{1},
			  V. Aynutdinov\IEEEauthorrefmark{1},
                          V. Balkanov\IEEEauthorrefmark{1},
                          I. Belolaptikov\IEEEauthorrefmark{4},
			  D. Bogorodsky\IEEEauthorrefmark{2},
                          N. Budnev\IEEEauthorrefmark{2},\\
                          I. Danilchenko\IEEEauthorrefmark{1},
                          G. Domogatsky\IEEEauthorrefmark{1},
			  A. Doroshenko\IEEEauthorrefmark{1},
                          A. Dyachok\IEEEauthorrefmark{2},
			  Zh.-A. Dzhilkibaev\IEEEauthorrefmark{1},\\
                          S. Fialkovsky\IEEEauthorrefmark{6},
			  O. Gaponenko\IEEEauthorrefmark{1},
                          K. Golubkov\IEEEauthorrefmark{4},
                          O. Gress\IEEEauthorrefmark{2},
			  T. Gress\IEEEauthorrefmark{2},
                          O. Grishin\IEEEauthorrefmark{2},\\
			  A. Klabukov\IEEEauthorrefmark{1},
                          A. Klimov\IEEEauthorrefmark{8},
                          A. Kochanov\IEEEauthorrefmark{2},
                          K. Konischev\IEEEauthorrefmark{4},
                          A. Koshechkin\IEEEauthorrefmark{1},
                          V. Kulepov\IEEEauthorrefmark{6},\\
                          D. Kuleshov\IEEEauthorrefmark{1},
                          L. Kuzmichev\IEEEauthorrefmark{3},
                          V. Lyashuk\IEEEauthorrefmark{1},
                          E. Middell\IEEEauthorrefmark{5},
                          S. Mikheyev\IEEEauthorrefmark{1},
                          M. Milenin\IEEEauthorrefmark{6},\\
                          R. Mirgazov\IEEEauthorrefmark{2},
                          E. Osipova\IEEEauthorrefmark{3},
                          G. Pan'kov\IEEEauthorrefmark{2},
                          L. Pan'kov\IEEEauthorrefmark{2},
                          A. Panfilov\IEEEauthorrefmark{1},
                          D. Petukhov\IEEEauthorrefmark{1},\\
                          E. Pliskovsky\IEEEauthorrefmark{4},
                          P. Pokhil\IEEEauthorrefmark{1},
                          V. Poleschuk\IEEEauthorrefmark{1},
                          E. Popova\IEEEauthorrefmark{3},
                          V. Prosin\IEEEauthorrefmark{3},
                          M. Rozanov\IEEEauthorrefmark{7},\\
                          V. Rubtzov\IEEEauthorrefmark{2},
                          A. Sheifler\IEEEauthorrefmark{1},
                          A. Shirokov\IEEEauthorrefmark{3},
                          B. Shoibonov\IEEEauthorrefmark{4},
			  Ch. Spiering\IEEEauthorrefmark{5},
			  O. Suvorova\IEEEauthorrefmark{1},\\
			  B. Tarashansky\IEEEauthorrefmark{2},
			  R. Wischnewski\IEEEauthorrefmark{5},
			  I. Yashin\IEEEauthorrefmark{3},
			  V. Zhukov\IEEEauthorrefmark{1}}
                            \\
\IEEEauthorblockA{\IEEEauthorrefmark{1}Institute for Nuclear Research of Russian Academy of Sciences,\\
     117312, Moscow, 60-th October Anniversary pr. 7a, Russia}
\IEEEauthorblockA{\IEEEauthorrefmark{2}Irkutsk State University, Irkutsk, Russia}
\IEEEauthorblockA{\IEEEauthorrefmark{3}Skobeltsyn Instutute of Nuclear Physics MSU, Moscow, Russia}
\IEEEauthorblockA{\IEEEauthorrefmark{4}Joint Institute for Nuclear Research, Dubna, Russia}
\IEEEauthorblockA{\IEEEauthorrefmark{5}DESY, Zeuthen, Germany}
\IEEEauthorblockA{\IEEEauthorrefmark{6}Nizhni Novgorod State Technical University, Nizhnij Novgorod, Russia}
\IEEEauthorblockA{\IEEEauthorrefmark{7}St.Petersburg State Marine University, St.Petersburg, Russia}
\IEEEauthorblockA{\IEEEauthorrefmark{8}Kurchatov Institute, Moscow, Russia}}

% please write the preseter's name and short title (3-4 words maximum)
%    which will appear at the header of the even pages.
\shorttitle{Author \etal paper short title}
\maketitle

\begin{abstract}
We present an analysis of neutrinos detected 
with the Baikal neutrino telescope NT200 for 
correlations with gamma-ray bursts (GRB). No 
neutrino events correlated with GRB were observed. 
Assuming a Waxman-Bahcall spectrum, a neutrino 
flux upper limit of {\bf $E^2 \Phi < 1.1 \times 10^{-6}cm^{-2}s^{-1}sr^{-1}GeV$ }
was obtained. We also present the Green's Function 
fluence limit for this search, which extends two orders 
of magnitude beyond the energy range of the Super-Kamiokande limit.
\end{abstract}
%====================================================

%====================================================
\section{Introduction}
The Baikal neutrino telescope NT200
\cite{1,2} is operating in Lake Baikal, Siberia, at a depth 1.1 km
since April, 1998. NT200 consists of 8 strings of 70 m length: 7 
peripheral strings and a central one. Interstring distances 
are about 20 m. Each string includes 24 pairwise 
arranged optical modules (OM). Each OM contains a 
37-cm diameter hybrid photodetector QUASAR-370. 

A number of relevant physics results has been obtained
so far with the NT200 telescope, e.g. limits on the diffuse
flux of extraterrestrial high energy neutrinos, limits on
neutrino fluxes from Dark Matter annihilation (Sun, Earth),
and on the flux of relativistic and slow magnetic monopoles
\cite{2,3,4}.

This work is devoted to the search of neutrino events 
correlated with observations of more than 300 
gamma-ray bursts (GRBs) reported from 1998 to 2000 
by the Burst and Transient Source Experiment (BATSE) \cite{5}.

The detection strategy for neutrino events with the 
NT200 telescope is based on a search for Cherenkov 
light from relativistic up-going muons produced by 
neutrino interactions.  Information about the GRB time 
and location on the sky allows to reduce the 
atmospheric muon background and, as a result, 
significantly increases the sensitivity of the neutrino 
telescope to neutrino events correlated with GRB.

%====================================================
\section{Experimental data}
\label{sect_ED}
For the present analysis, the experimental data obtained 
with NT200  from April 1998 to May 2000 were used. The selected
data sample contains those events which were formally reconstructed
as up-going muons. Taking into account the high level of background 
for directions close to the horizon, only events with zenith angles
larger than $100^\circ$ were selected. The average rate of such 
events was 0.037 Hz. Most fake events are due to misreconstructed 
muons close to horizon and to muon bundles.  

For the present analysis of time and directional correlations 
with NT200 events we used information about GRB location, time,
duration $T_{90}$, and location error from the basic BATSE 4B catalog \cite{5}
(triggered bursts) and from the catalog of non-triggered GRB \cite{6}.
The error distribution of BATSE GRB locations was taken from \cite{7}.
A total of 303 GRBs (155 triggered and 148 non-triggered) at zenith angles
larger than $100^\circ$  and occurring during periods of stable 
operation of NT200 have been selected.

%================================================================================
\section{Data selection criteria and detector effective area}
\label{sect_DS}
The optimization of the data selection criteria was performed on the 
basis of simulated neutrino events \cite{8} and events of atmospheric
 muon background \cite{9} in NT200. Taking into account the varying
 NT200 configurations during the considered time period, calculations
 have been performed for nine basic detector configurations, most of
 them closely corresponding to the real status of the detector. 

The results of the reconstruction of simulated events were used to 
estimate the reconstruction efficiency and to calculate
the background. Event reconstruction and data selection for NT200 
are described in detail in \cite{8}. There, selection criteria 
were designed and optimized for atmospheric neutrino separation. They provide a 
rejection factor of atmospheric muons larger than $10^7$. For GRB, the aditional 
information about detection time and location on sky, however, 
allows softening the requirements to the background rejection.
This increases the registration probability for useful events and 
therefore greatly increases the effective neutrino detection area. 
Following the approach of \cite{8},
$P_{hit} \times P_{nohit}$ and $Z_{dist}$ were chosen as basic parameters
for event selection. $Z_{dist}$ is the maximal distance between
all projections of the triggered OM coordinates onto the reconstructed 
muon trajectory. $P_{hit}$ is the normalized probability 
of fired channels to be hit, and $P_{nohit}$ is the probability of 
non-fired channel not to be hit.

For the present correlation analysis, two sets of criteria for event 
selection were chosen:
\\Cut-A:
\\$(Z_{dist}>30m)\&(P_{hit}\times P_{nohit}>0.1)\&(\Psi<10^\circ)$,
\\Cut-B:
\\$(Z_{dist}>30m)\&(\Psi<5^\circ)$,
\\were $\Psi$ is the angle between up-going muon and GRB-direction. 

Cut-A dominantly selects neutrinos with energies below $\sim10^6$ GeV.
Cut-B allows a significant extension of the energy range, but the 
expected background is approximately four times larger than for Cut-A.

Calculating the effective area of NT200  for the two sets of criteria, 
we took into account the absorption of neutrinos passing through the 
Earth, as well as the production, propagation, detection and 
reconstruction within a given angular cut $\Psi$ of muons. The 
calculated effective areas for the Cut-A and Cut-B samples are presented 
in Fig. 1 as a function of neutrino energy.

The effective areas for the two sets are close to each other up to
$\sim10^5$ GeV. For larger energies, the effective area for Cut-A 
stays essentially constant. The behavior for $E>10^6$ GeV is largely 
defined by neutrino absorption in the Earth.

The energy range of the NT200 sensitivity was estimated for an $E^{-2}$ 
neutrino spectrum. The 90\% sensitivity range of NT200 
extends up to $\sim10^6$ GeV and $\sim10^7$ GeV for selection criterion 
A and B, respectively.

\begin{figure}[htb]
\centering
\includegraphics[width=0.50\textwidth]{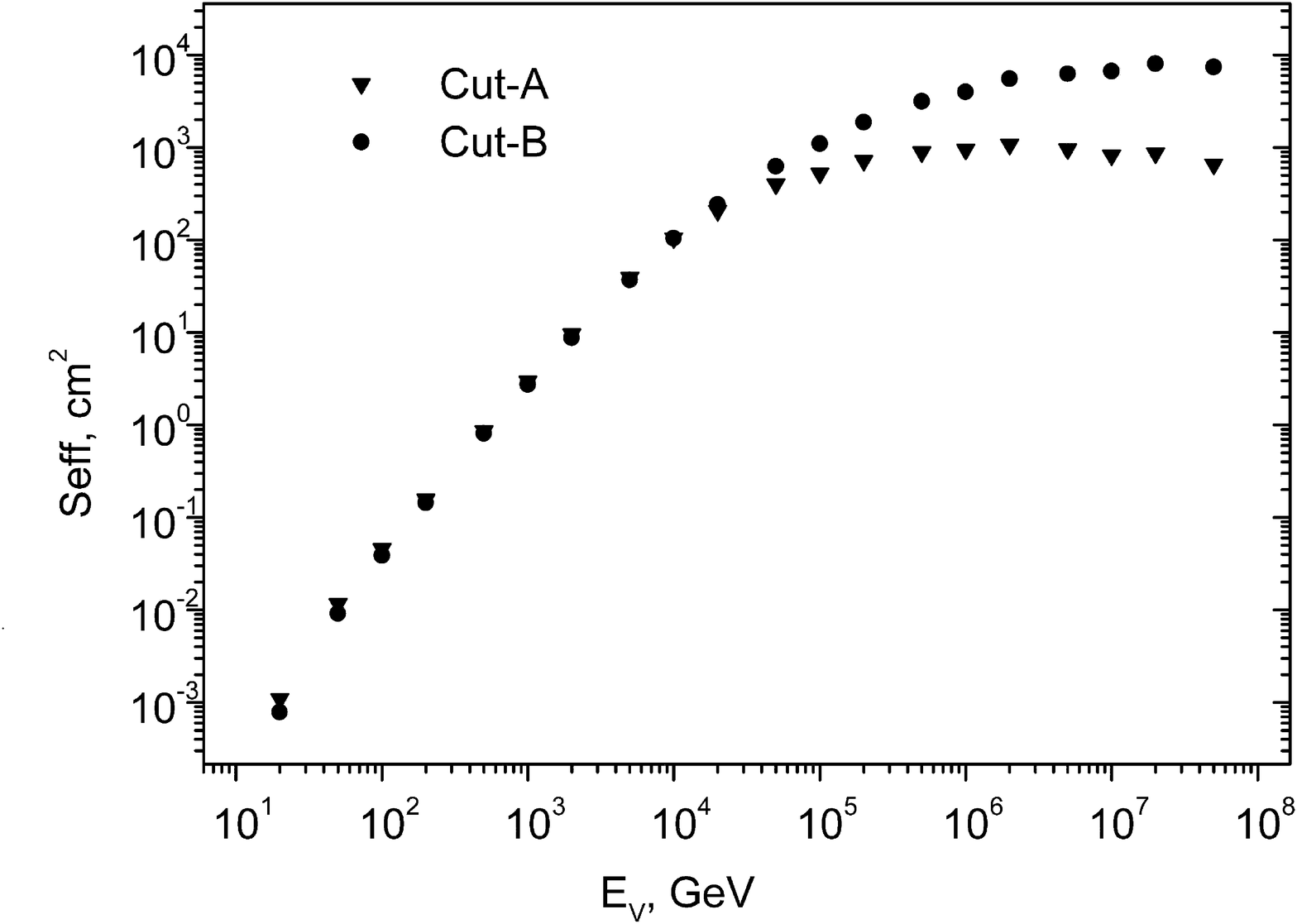}
\caption{NT200 effective area averaged over zenith angles 
between $100^\circ$ and $180^\circ$ as a function of neutrino energy, for selection criteria Cut-A and Cut-B.}
\label{fig_01}
\end{figure}

A global estimation of the background, expected for the GRB search time window,
was obtained from the NT200 raw event rate (0.037 Hz), the calculated atmospheric
muon rejection factor for Cut-A and Cut-B, and the total GRB duration
 $T_{GRB}$ ($\sim1.8 \times 10^4$ s). $T_{GRB}$ was calculated as the
 sum of the $T_{90}$ intervals for all 303 triggered and non-triggered GRB.
 For compensation of possible event time uncertainties, five seconds were 
 added at both sides of the time intervals $T_{90}$. For cases of missing
 information on $T_{90}$ (about 25\% of the triggered GRB), a fixed time
 interval was used. The expected background values, obtained from the full
 detector livetime interval, are 0.8 and  3.1 events for Cut-A and Cut-B,
 respectively.  
 
%=================================================================
\section{Data analysis and results}
\label{sect_DA}
The basic objective of this study was to verify a sufficiently
precise detector responce simulation, to search for events
correlated with GRB, and to provide a solid background estimation. 

To check the simulation procedures, the calculated atmospheric muon 
rejection factors were compared to experimental values. The results
are presented in Table I for different criteria
 $P_{hit}\times P_{nohit}$ ($Z_{dist}$ is set to 30 m). 
 The simulated values are in agreement with the experimental results
 within the systematic error of our calculation, about 20\%.
 
\begin{table}[!h]
  \caption{Atmospheric Muon Rejection Factor: Experiment and Simulation}
  \label{table_1}
  \centering
  \begin{tabular}{|c|c|c|c|}
  \hline
   $P_{hit}\times P_{nohit}$  & $\geq0.1$ & $\geq0.2$ & $\geq0.3$\\
   \hline 
                     &       &       &        \\
    {\it Experiment} & 0.053 & 0.012 & 0.0035 \\
                     &       &       &        \\
    {\it Model}      & 0.062 & 0.014 & 0.0040 \\
  \hline
  \end{tabular}
  \end{table}

This search for correlation with NT200 neutrino events uses
303 GRBs (triggered and non-triggered) selected from the total sample 
of 736 BATSE GRBs recorded in 1998-2000. The search for events selected 
according to criteria A and B was done for the time intervals $T_{GRB}$
(see section III). For a detailed background estimation, we used a time interval
$\pm1000$ s with respect to the start time of the GRB (excluding the
 $T_{GRB}$ window) and an angle between GRB and up-going muon 
 $\Psi \leq 10^\circ$.
 
No events were found according to criterion A and one event by
 criterion B. Table II shows the number of signal and background events,
 the event upper limit $\mu_{90}$  that was obtained in accordance with [10],
 the number of GRB corrected to coefficient $\beta$ (the probability
 to detect events associated with gamma-ray burst in the given angle interval
 $\Psi$, given the location error of the GRB), and the 90\% C.L. upper
 limit on the number of events per GRB 
 $N_{90}=\mu_{90}$ / $(N_{GRB}\times \beta)$. 
 Data are presented for all GRB (triggered and non-triggered), as well as
 separatly for triggered GRB selected according to Cut-B. 

\begin{table}[!h]
  \caption{Results of GRB Analysis}
  \label{table_2}
  \centering
  \begin{tabular}{|l|c|c|c|c|c|}
  \hline
   {\it Selection}  & {\it Signal} & {\it Backgr} & $\mu_{90}$ & $N_{GRB} \times \beta$ & $N_{90}$ \\
   \hline 
                      &       &       &       &      &        \\
    {\it Cut-A, all}  &  0    & 0.56  & 1.9   & 236  & 0.0085 \\
                      &       &       &       &      &        \\
    {\it Cut-B, all}  &  1    & 2.7   & 2.1   & 199  & 0.010  \\
                      &       &       &       &      &        \\ 
    {\it Cut-B, trig} &  1    & 1.6   & 2.8   & 120  & 0.023  \\    
  \hline
  \end{tabular}
  \end{table}

No excess of events associated with a GRB was observed. The limit 
on the neutrino flux associated with gamma-ray bursts was 
obtained using the approach from \cite{11}.
According to this approach, the limit $F(E_{\nu})$ is presented as 
a function on neutrino energy, the "Green's function":
\begin{equation}
F(E_{\nu})=N_{90} / S_{eff}(E_{\nu}),
\end{equation}
$S_{eff}(E_{\nu})$ is the detector effective area, 
$N_{90}$ the 90\% C.L. upper limit on number of events per GRB. 
The main advantage of this approach is that its result does not 
depend on assumptions about the neutrino energy spectrum.

\begin{figure}[htb]
\centering
\includegraphics[width=0.50\textwidth]{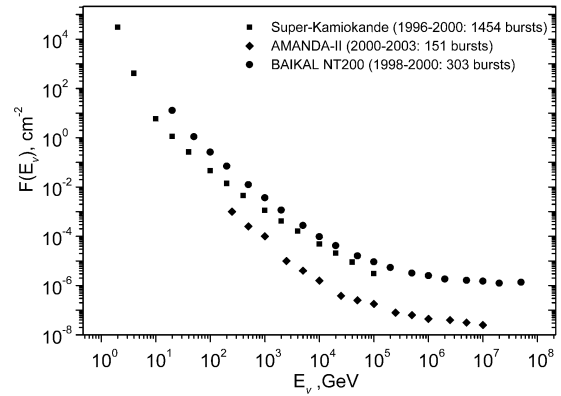}
\caption{$90\%$ C.L. upper limits on the GRB neutrino fluence 
Green's function $F(E_{\nu})$ for NT200, Super-Kamiokande  and AMANDA.}
\label{fig_02}
\end{figure}

Figure 2 shows the 90\% C.L. upper limits of the GRB neutrino fluence
Green's function $F(E_{\nu})$ for NT200 (for Cut-B, all), Super-Kamiokande 
\cite{11} and AMANDA \cite{12}. 
The Super-Kamiokande and NT200 limits are mainly for GRB 
from the southern sky, while the AMANDA limit is for the northern sky.

Since predictions for the energy spectrum of neutrinos from GRB
differ from model to model, we prefer to present our basic
experimental result as a Green's function $F(E_{\nu})$,  
which allows to calculate limits for any neutrino energy
spectrum. In addition, we translate our result to a benchmark spectrum.
We have chosen that of E.Waxman and J.Bahcall \cite{14}, 
and E.Waxman \cite{15,16}. Following these works, the muon neutrino
 differential flux $\Phi_{W-B}(E_{\nu})$  in the energy range up to 10 PeV is:
\begin{equation}
E_{\nu}^{2}\,\Phi_{\nu}^{W-B}(E_{\nu})=A^{W-B}\times \min(1,E_{\nu}/E_{\nu b}),
\end{equation} 
$E_{\nu b}$=100 TeV, 
$A_{W-B}\approx 8\times 10^{-9}GeVcm^{-2}s^{-1}sr^{-1}$. 

The Model Rejection Factor MRF for the Waxman-Bahcall spectrum was
calculated as: 
\begin{equation}
MRF=N_{90} / N_{ex},
\end{equation} 
where $N_{90}$ is the upper limit on the number of events per GRB and
$N_{ex}$ the  expected number of events, calculated for the
given spectrum as:
\begin{equation}
   N_{ex}=\int \Phi_{\nu}^{Earth}(E_{\nu})
   \times S_{eff} (E_{\nu})\times(4\pi/n)dE_{\nu}.
\end{equation}
Here, $n \approx 2.2 \times 10^{-5} s^{-1}$ is the average GRB
rate in $4{\pi}sr$ ($\sim700$ events within the detection range
of the BATSE per year) and $\Phi_{\nu}^{Earth}(E_{\nu})$=$0.5\times\Phi_{\nu}^{W-B}(E_{\nu})$
the neutrino flux at the Earth.

Taking into account that the estimation of the expected event number
in the given approach is made for for the BATSE burst detection rate,
the MRF was calculated only for triggered GRB ($N_{GRB} \times \beta =120$,
see Table II). From that Green's function (approximately twice that 
of the (Cut-B, all) sample, see Fig.2), the resulting MRF value is
$2.8 \times 10^2$, and the corresponding GRB neutrino flux limit is

\begin{equation}
E_{\nu}^2\,\Phi_{\nu} \leq 1.1 \times 10^{-6} GeVcm^{-2}s^{-1}sr^{-1}  
\end{equation}    

This diffuse limit is considerably weaker that of the AMANDA
muon analysis \cite{12}, and two times higher that their cascade
analysis \cite{13}. In view of a search for bright individual GRBs,
our result may be considered as complementary to AMANDA
since variations in absolute energy output, Lorentz factor 
and distance might lead to a GRB neutrino detection with a less
sensitive detector, while that source was outside the other detector
field of view. NT200+ is presently complementing the ANTARES detector,
in particular for those cases where the GRB is above horizon with respect
to this instrument. We also not that, normalized to a single GRB,
NT200 exceeds the sensitivity of Super-Kamiokande by a factor
of 2 for neutrino energy above 1 TeV.

\section{Conclusion}

We have presented results of a search for neutrino induced muons 
detected with Baikal Telescope NT200 in coincidence with 
303 gamma-ray bursts recorded from 1998 to 2000 by BATSE. 
NT200's field of view covers most part of the Southern hemisphere.
No evidence for neutrino-induced  muons from gamma-ray bursts is found.
The resulting Green's Function fluence limit for this search extends 
that of Super-Kamiokande by two orders of magnitude in energy.
Assuming a Waxman-Bahcall spectrum, a neutrino flux upper limit of
% \begin{equation}
\\$E_{\nu}^2 \, \Phi_{\nu} \leq 1.1 \times 10^{-6} GeVcm^{-2}s^{-1}sr^{-1}$  
% \end{equation} 
  is obtained.

\section{Acknowledgments}
%% Keep the small font tag for the acknowledgments
{\small

This work was supported in part by the Russian Ministry of Education
and Science, by the German Ministry of Education and Research,
by the Russian Found for Basic Research (grants 08-02-00432-a,
07-02-00791, 08-02-00198, 09-02-10001-k, 09-02-00623-a), by the grant of
the President of Russia NSh-321.2008-2 and
by the program "Development of Scientific Potential in Higher Schools"
(projects 2.2.1.1/1483, 2.1.1/1539, 2.2.1.1/5901).
}

%%%%%%%%%%%%%%%%%%%%%%%%%%%%%%%%%%%%%%%%%%%%%%%%%%% 

\end{document}